# Structural Determination of Nanocrystalline Si Films Using Ellipsometry and Raman Spectroscopy

Sanjay K. Ram[a*], Md. N. Islam[b], P. Roca i Cabarrocas[c] and Satyendra Kumar[a†]

[a]*Department of Physics & Samtel Center for Display Technology, Indian Institute of Technology Kanpur, Kanpur-208016, India*
[b] *QAED-SRG, Space Application Centre (ISRO), Ahmedabad – 380015, India*
[c]*LPICM, UMR 7647-CNRS, Ecole Polytechnique, 91128 Palaiseau Cedex, France*

Single phase nano and micro crystalline silicon films deposited using $SiF_4$/ $H_2$ plasma at different $H_2$ dilution levels were studied at initial and terminal stages of film growth with spectroscopic ellipsometry (SE), Raman scattering (RS) and atomic force microscopy (AFM). The analysis of data obtained from SE elucidates the microstructural evolution with film growth in terms of the changes in crystallite sizes and their volume fractions, crystallite conglomeration and film morphology. The effect of $H_2$ dilution on film microstructure and morphology, and the corroborative findings from AFM studies are discussed. Our SE results evince two distinct mean sizes of crystallites in the material after a certain stage of film growth. The analysis of Raman scattering data for such films has been done using a bimodal size distribution of crystallite grains, which yields more accurate and physically rational microstructural picture of the material.

PACS numbers: 68.55.-a, 68.55.Jk, 68.35.Ct

Keywords: Silicon; Thin films; Structural properties; Growth mechanism; Ellipsometry; Atomic force microscopy (AFM); Raman spectroscopy.

## 1. INTRODUCTION

Recent demonstration of thin film solar cells produced fully from nanocrystalline (or microcrystalline) silicon has generated a new wave of interest in undoped nanocrystalline silicon as this material offers increased stability against light induced degradation and has enhanced low-energy absorption in contrast to amorphous silicon (*a*-Si:H) [1,2,3]. Hydrogenated nanocrystalline silicon (nc-Si:H) is also attractive for thin film transistors and other photovoltaic devices due to its easy fabrication technology allowing large area deposition even at low temperatures, on non-crystalline substrates [4,5]. The stability and efficiency of the nc-Si:H films are important issues for these device applications and both these factors are determined by the microstructural characteristics of the film [6,7]. The microstructural characterization of the component layers are thus vital to the understanding and prediction of the device efficiency and properties. The structural evolution of plasma deposited nc-Si:H films can be manipulated to yield maximum efficiency by altering the deposition parameters, mainly, hydrogen dilution. The knowledge of film microstructure at various levels of film thicknesses and its relationship with different $H_2$ dilutions provides a method to predict and assess the outcome of the controlled deposition techniques.

Microstructurally, plasma deposited nc-Si:H and microcrystalline silicon ($\mu$c-Si:H) are heterogeneous materials consisting of Si nanocrystallites ($\approx$ 2 to 20 nm) and aggregates thereof (which may reach up to few hundred nm in $\mu$c-Si:H), dispersed in an amorphous Si matrix, and disorder in the form of voids and grain boundaries. In the absence of an amorphous phase in this material (which is then termed as single-phase nc-Si:H or $\mu$c-Si:H), the disordered phase consists of voids, density deficit and clusters of nanocrystallites. The heterogeneous nature of nc-Si:H microstructure exists along the timescale of growth as well, and with an increase in film thickness, a distribution in the sizes of crystallite grains can be observed, with variation in the fractional compositions of the different sized constituent grains.

Spectroscopic ellipsometry (SE) and Raman spectroscopy (RS) are basic non-destructive optical tools routinely employed to study the microstructural properties of nc-Si:H films. In situ SE provides important information about the film thickness, delineates film layers on the basis of their components and identifies the volume fractions of these components as well. RS provides valuable information regarding the film crystallinity and the nature of crystalline composition. It is apparent, that the information derived from the SE and RS study depends on the relevance and accuracy of the data analysis method. The conventional approach to the deconvolution of Raman spectra is the same for films of any growth stage, and does not take into account the presence of the size distribution [8]. This may result in some significant errors in the calculation of total crystallinity (may be underestimated) and the mean crystallite size (may be overestimated) [9].

[*] corresponding author. E-mail: skram@iitk.ac.in , sanjayk.ram@gmail.com
[†] satyen@iitk.ac.in

In this paper we present a microstructural study of single-phase nc-Si:H thin films at the very early and terminal stages of film growth, deposited at different $H_2$ dilutions. Our contribution aims to bring out the quantitative microstructural differences between films at different stages of growth, and the influence of $H_2$ dilution on the film microstructure and morphology. We propose a need for differential approach to microstructural data analysis based on the stage of film growth. Our results evidence a significant presence of size distribution in crystallite grains (CSD) in films, which varies with the stage of film growth. We have presented an approach to the deconvolution of Raman spectra of nc-Si:H based on a model incorporating the CSD [9].

## 2. EXPERIMENTAL DETAILS

The undoped nc-Si:H films were deposited at low substrate temperature (≤ 200˚C) in a parallel-plate glow discharge plasma enhanced chemical vapor deposition system operating at a standard rf frequency of 13.56 MHz, using high purity $SiF_4$, Ar and $H_2$ as feed gases. Different microstructural series of a large number of samples were created by systematically varying gas flow ratios ($R$ = $SiF_4/H_2$) for samples having different thicknesses ($d \approx$ 50-1200nm). Each nc-Si:H sample was studied with different microstructural tools like RS, in-situ phase modulated SE, X-ray diffraction (XRD), and atomic force microscopy (AFM) to obtain comprehensive and consistent microstructural information [10]. Bifacial Raman scattering measurements were carried out by focusing the light probe (He-Ne laser, ≈632.8nm) from the top film surface [denoted here as RS(F)], as well as from the bottom surface [through the glass substrate, denoted here as RS(G)] on the samples to obtain information about the film at different stages of growth. In this report, the samples belong to two sets, the first set has films with thickness in the range of 55 nm, and the second set has films having thickness ≈ 950 nm, to ensure comparability of the growth stage between the samples of a particular set.

## 3. RESULTS AND DISCUSSION

An important indication of the nucleation processes occurring at the initial stage of growth, and a fair prediction of the microstructural growth at higher film thicknesses can be obtained from the in-situ spectroscopic study of films at very early stages of growth. Fig. 1(a) shows the imaginary part of the pseudo-dielectric function $<\varepsilon_2>$ spectra measured by SE on such samples at initial stages of growth (average $d \approx$ 55 nm), deposited under different $R$ values. Energy positions at ≈ 3.4 eV ($E_1$) and at ≈ 4.2 eV ($E_2$) are marked in this graph to denote the two prominent structures of the optical absorption in c-Si, well predicted by the theoretical electronic band structure, and particularly sensitive to imperfect crystallinity. In contrast, such peaks are not seen in the $<\varepsilon_2>$ spectrum of *a*-Si:H, rather a broad distribution with a maximum at around $E_1$ is seen. In the case of poly or nano or microcrystalline silicon, broad shoulders at the energy positions $E_1$ and $E_2$ are seen. The $E_2$ peak in Fig. 1 is completely absent from the spectra of the film deposited at $R$ = 1/1, while a faint indication of such a peak can be seen in the other two films ($R$ = 1/10 and 1/20). Thus, with increasing $H_2$ dilution, there is an improvement in film crystallinity. High $H_2$ dilution is known to etch out the weak and unsaturated bonds more effectively, resulting in a better crystalline material. The amplitude of the $<\varepsilon_2>$ spectrum of the $R$= 1/20 film is very low compared to the other two films because such a high $H_2$ dilution is known to result in a formation of larger crystallites or crystallite aggregates, with an associated increase in the amount of voids, thereby decreasing the overall crystalline volume fraction.

Experimental SE data was fitted using standard Bruggeman effective medium approximation. A well-chosen reference crystalline silicon (c-Si) is essential for obtaining a good fit and a physically rational description of the film structure. We have used published dielectric functions for low-pressure chemical vapor deposited polysilicon with large (pc-Si-l) and fine (pc-Si-f) grains [11]. The schematic view of the microstructures of the same films obtained from modeling of the SE data is presented in Fig. 1(b), where the fractional composition of constituent components and the layer thicknesses (**d**$_l$) are mentioned. The percentage volume fractions of fine grains, large grains, voids and amorphous content are denoted by $F_{cf}$, $F_{c1}$, $F_v$ and $F_a$ respectively. In this model diagram, we can see that with rise in $H_2$ dilution amorphous content in the bulk layer (BL) decreases, resulting in a material totally free of any amorphous content at the dilution level of $R = 1/20$. However, the initial increase in $F_{cf}$ in the BL seen on increasing the $R$ from 1/1 to 1/10, denoting improved crystallization, drops sharply with a further increase in $R$ (1/20), while a simultaneous increase in $F_v$ takes place. The $R = 1/20$ film shows increased void fraction due to the known effect of high $H_2$ dilution that results in less nucleation site generation and increased etching.

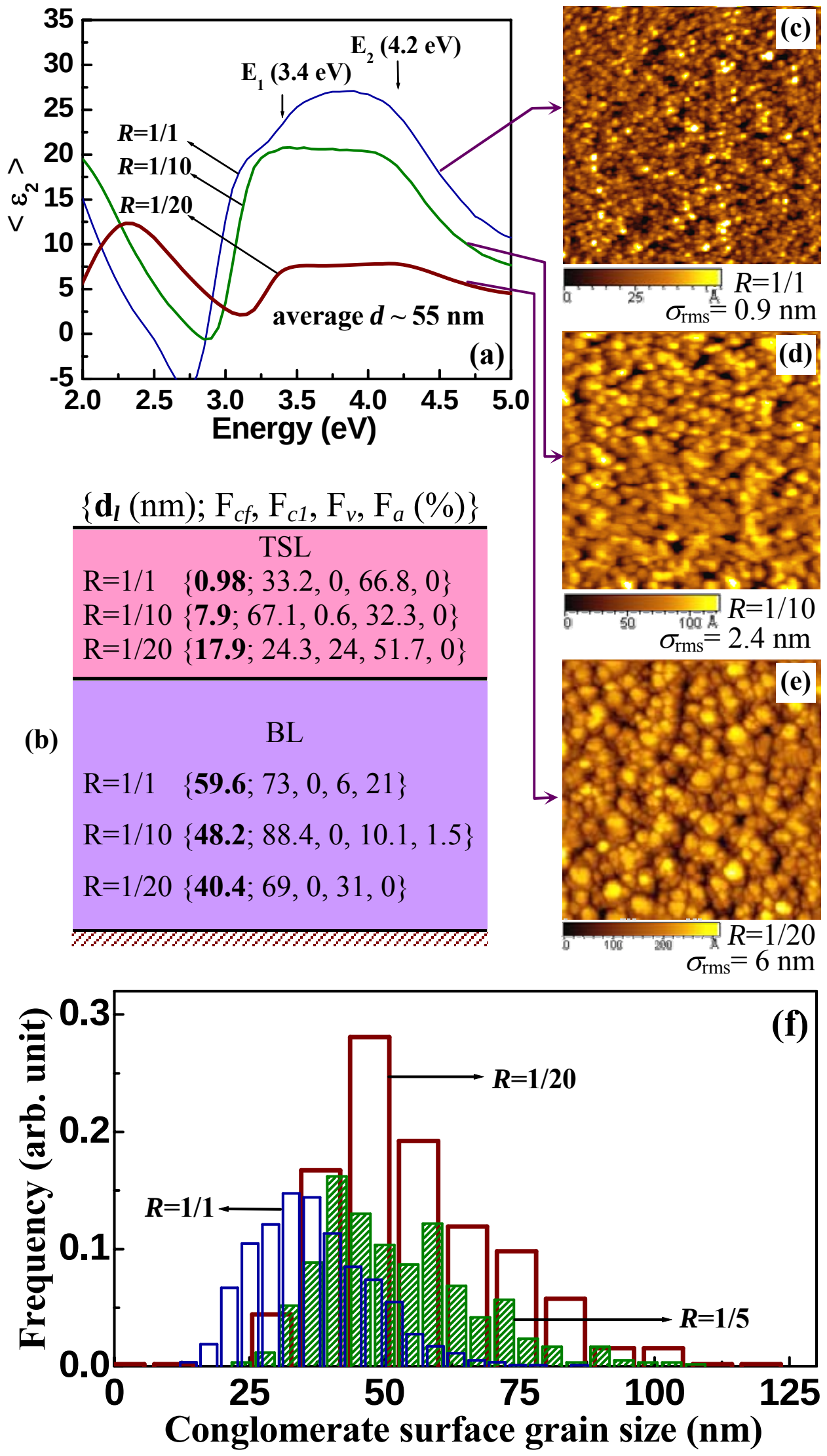

FIG. 1. (a) Measured imaginary part of $<\varepsilon_2>$ spectra for nc-Si:H samples of early stage of growth ($d \approx$ 55 nm) prepared by different $R$ values (1/1, 1/10 and 1/20); (b) schematic illustration of microstructures of the same films obtained from SE data analysis; the parts (c), (d) and (e) show the 2D-AFM images (1000 nm × 1000 nm) of the same films; (f) the histograms of size distribution of conglomerate surface grains in these films.

The top surface layer (TSL) in Fig. 1(b) demonstrates well the influence of $H_2$ dilution on the microstructural remodeling of the nc-Si:H films. The film deposited at $R$= 1/1 has the least top surface roughness layer thickness and contains only small crystallite grains. With an increase in the $H_2$ dilution, a continuous rise in the roughness layer is observed, which reaches the highest value for the highest $H_2$ dilution case. Simultaneously, higher $H_2$ dilution favors the formation of large crystallite grains, and large grains make an appearance in the films deposited under $R$= 1/20. These findings of surface layer are substantiated by the results of AFM study as well. The 2-D AFM images of the same three films are presented in Fig. 1(c), (d) and (e). Here, we see that on increasing the $H_2$ dilution, the size of surface grains increases [Fig. 1(d) and (e)]. These films having $d \approx 55$ nm can be assumed to denote the beginning of the growth process, here the grain density is found to be higher at a lower value of $H_2$ dilution. A corroborative increase in surface roughness is also seen with increasing $H_2$ dilution at this stage of film growth, as mentioned in each figure. The histograms of the size distribution of surface grains of these samples are shown in Fig. 1(f), where a shift in the distribution towards bigger size is observed as $H_2$ dilution is increased.

Applying such known considerations about the process of grain growth in plasma deposited nc-Si:H, one can reasonably predict the further growth patterns of these films from the microstructural picture observed at the early growth stage. Thus, the film deposited at lower $H_2$ dilution is expected to have tightly packed crystallites and conglomeration may assume the shape of straight columns. In contrast, for films deposited under higher $H_2$ dilution, more lateral growth of crystallites may be seen with the preservation of the individuality of the crystallites. Now let us verify these predictions by looking at the results of films at the terminal stages of growth. To explore how the film microstructure changes over the growth process, the imaginary part of $\langle\varepsilon_2\rangle$ spectra measured by SE on samples having average $d \approx 950$ nm, deposited under different $R$ values (1/1, 1/5 and 1/10), is presented in Fig. 2(a). These films have reached a steady state growth stage, and the associated changes in the bulk crystalline material with increase in film thickness is reflected in the increase in the intensity of the shoulder of $E_2$ peak, indicating excellent crystallization. The schematic model diagram of the microstructure of the same three films is shown in Fig. 2(b), where the microstructure is seen to have changed remarkably compared to the initial stages. At this stage, three layers of the film can be distinguished, and the modeling of SE data has been done using a three layer model: a top surface layer (TSL), a middle bulk layer (MBL) and a bulk interface layer (BIL).

In Fig. 2(b), with an increase in $H_2$ dilution, the incubation layer or BIL thickness is seen to decrease. An initial rise in $H_2$ dilution eliminates the amorphous content in the BIL with improvement in overall crystallinity, but a further rise to $R = 1/10$ results in a lowered crystalline fraction with an associated increase in $F_v$. The middle bulk layer is devoid of any amorphous content, having both small and large crystallite grains in varying ratios for the three films. The variation in the ratios of the crystallite grains of two different mean sizes with film growth that is recognized by our SE analyses in these films was identified in all our samples of other intermediate thicknesses also, which are not shown here. In Fig. 2(b), again a very high $H_2$ dilution level is seen to result in a higher void fraction in all the layers. The AFM images of these samples in Fig. 2 (c), (d) and (e) reveal that the con-glomeration of the grains has taken place, resulting in formation of large crystallite grains. The conglomerate grains of $R$=1/5 and $R$=1/1 films are found to cover a wider area by collecting more grains into their domains, compared to the sample of $R$=1/10. The film of $R$= 1/10 contains conglomerate grains having a cauliflower-like shape, which implies that the growth in lateral direction is more when the $H_2$ dilution is increased, which results in a bigger size of conglomerate grains. These observations are in good agreement with our earlier predictions. The surface roughness values mentioned in the respective AFM images indicate the same finding as in Fig. 2(b), i.e., the roughness layer increases with increasing $H_2$ dilution.

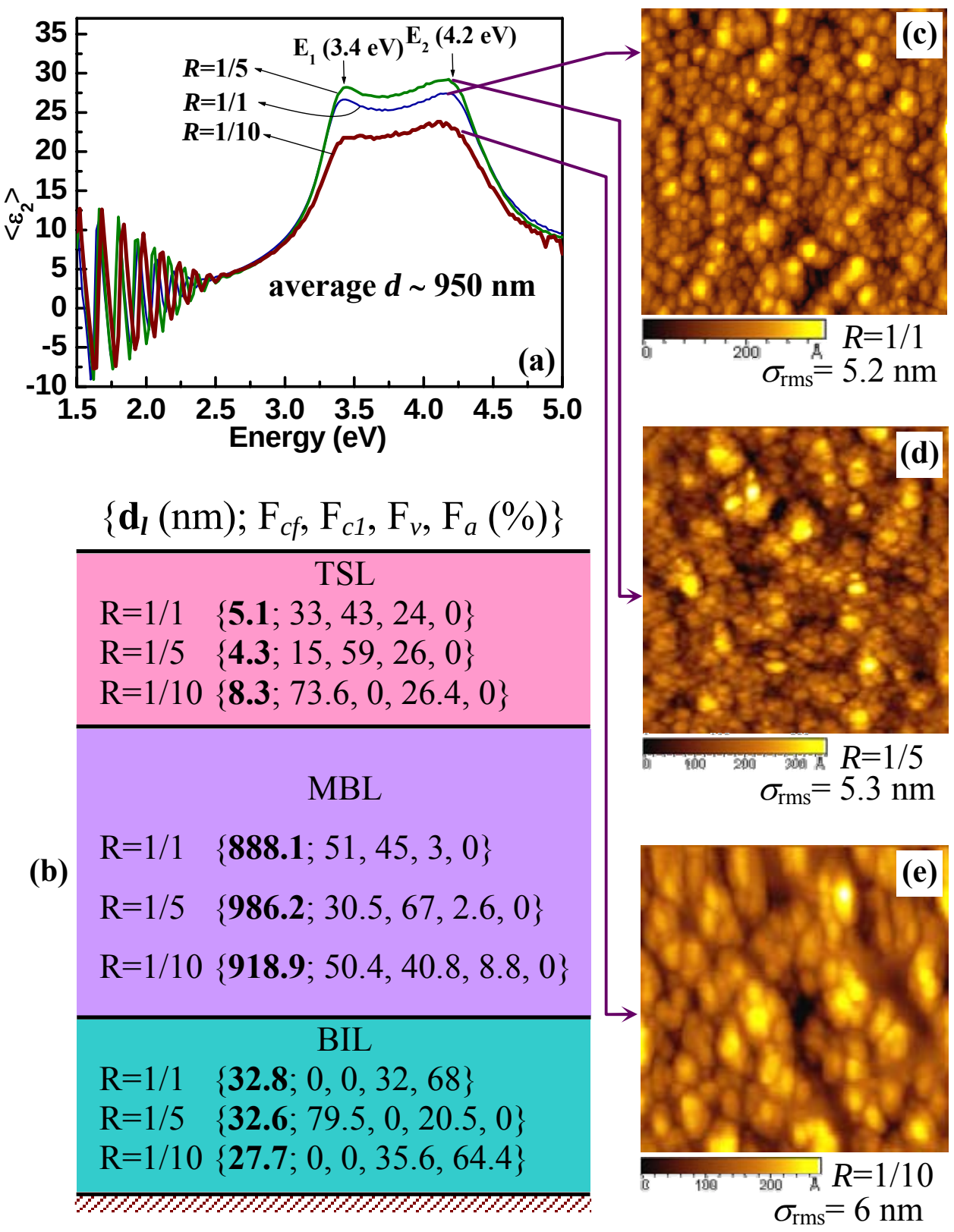

FIG. 2. (a) Measured imaginary part of $\langle\varepsilon_2\rangle$ spectra for nc-Si:H samples of terminal stage of growth ($d \approx 950$ nm) prepared by different $R$ values (1/1, 1/5 and 1/10); (b) schematic illustration of microstructures of the same films obtained from SE data analysis; the parts (c), (d) and (e) show the 2D-AFM images (2000 nm × 2000 nm) of the same films.

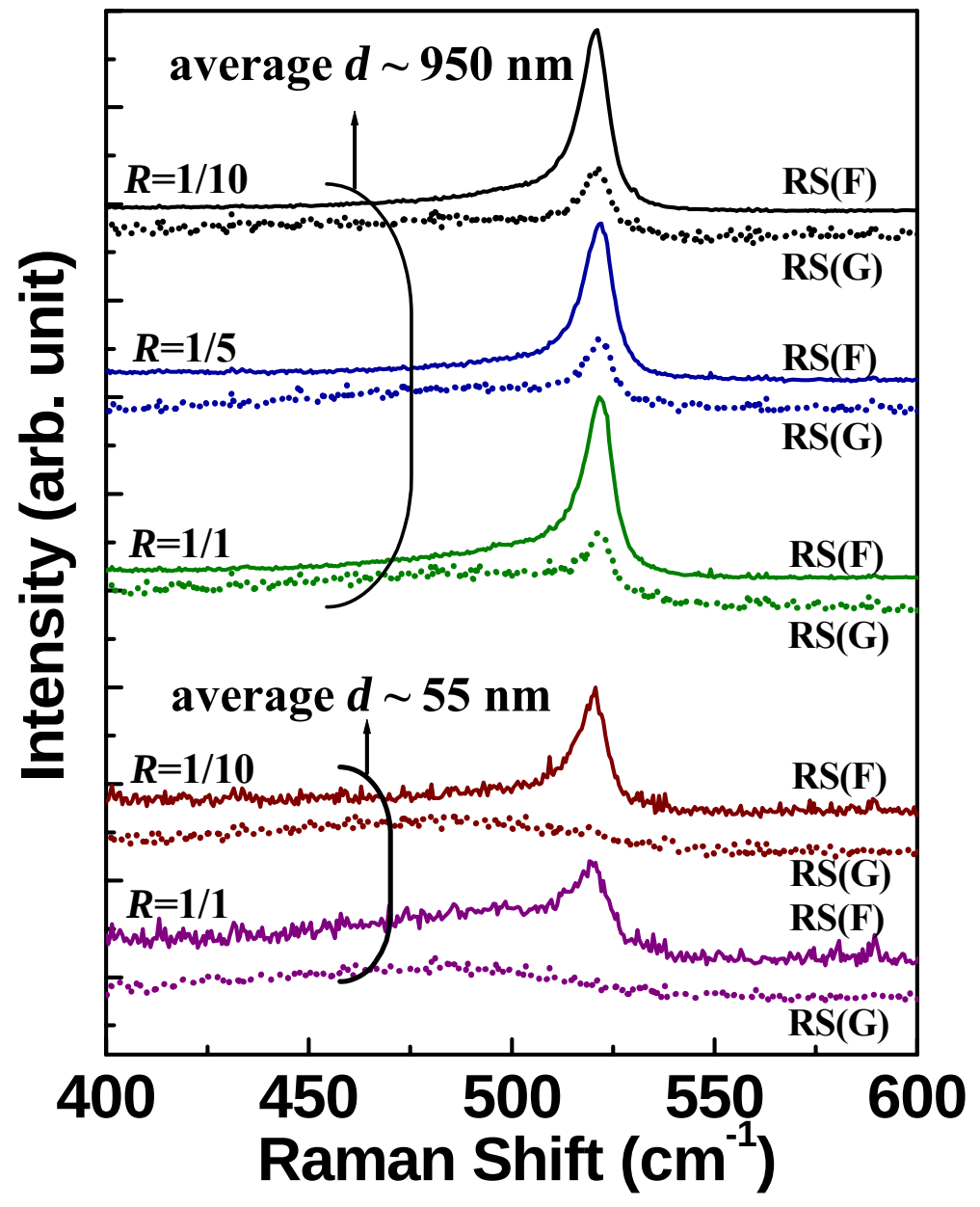

FIG. 3. Bifacial Raman spectra [RS(F) and RS(G)] for nc-Si:H samples of two stages of growth (at average $d \approx 55$ and 950 nm) and deposited under different $R$ values.

Now we come to the results of the bifacial RS measurements of nc-Si:H samples at different stages of growth (at average $d \approx 55$ and 950 nm) and deposited under different $R$ values. The RS(F) and RS(G) profiles of these samples are presented in Fig. 3. For the set of samples having $d \approx 950$, we see that an explicit amorphous hump (a broad peak at 480 $cm^{-1}$) is absent in their RS(F) profiles. The amorphous hump can be seen in the RS(G) of all the samples of both the thickness sets, due to the presence of amorphous content in the interfacial layer. All these findings are in good agreement with the SE results. However, in the samples at early stages of growth, such an amorphous hump can be seen in the RS(F) profile. In the absence of an explicit amorphous hump, the asymmetry in the Raman lineshape of RS profiles of nc-Si:H samples, seen as a low energy tail, can be attributed to the distribution of smaller sized crystallites.

The presence of two distinct sizes of crystallites and their varying percentage volume fractions at different stages of film growth deduced from SE results indicate a need for taking into consideration this CSD while deconvoluting the RS profiles, for a better agreement and physical accuracy. Keeping these considerations in mind, we have used the deconvolution model described in Ref. [9] for obtaining the CSD in Si nanostructures, for the analysis of our RS data. The final deconvolution results consisting of the total fit, and the deconvoluted peaks for the RS profiles of one samples from each thickness set are shown in Fig. 4. The RS(F) and RS(G) profiles of $R$ = 1/10 film having $d \approx 55$ nm are shown in Fig. 4(a) and (b) respectively, while the RS(F) and RS(G) profiles of $R$ = 1/5 film having $d \approx 950$ nm are shown in Fig. 4(c) and (d) respectively.

Taking the inputs from SE studies basically gives rise to three different models which can be applied to the deconvolution of the RS profiles of our nc-Si:H samples:

(i) Model “cd+a” having only one type of CSD (mainly the contribution from small sized crystallites) with an amorphous background: Fig. 4(a) and (b).

(ii) Model “cd1+cd2” where a bimodal CSD [small (cd1) and large (cd2) crystallites] without an amorphous background is considered: Fig. 4(c).

(iii) Model “cd1+cd2+a” incorporating a bimodal CSD along with an amorphous phase: Fig. 4 (d).

These fitting models were applied to all the RS profiles of the samples, and the best fit is shown in these graphs. For assessment of the goodness of fit, not only are the visual scrutiny and statistical validation important, but the physical plausibility, the corresponding SE data and the direction of RS measurement also needed to be taken into account. The direction of RS measurement is important because the Raman collection depth for $\mu$c-Si:H material at 632.8 nm (He-Ne red laser) is $\approx 500$ nm, which has to be kept in view, especially in the samples at lower thicknesses, to understand the data collected by RS from each direction.

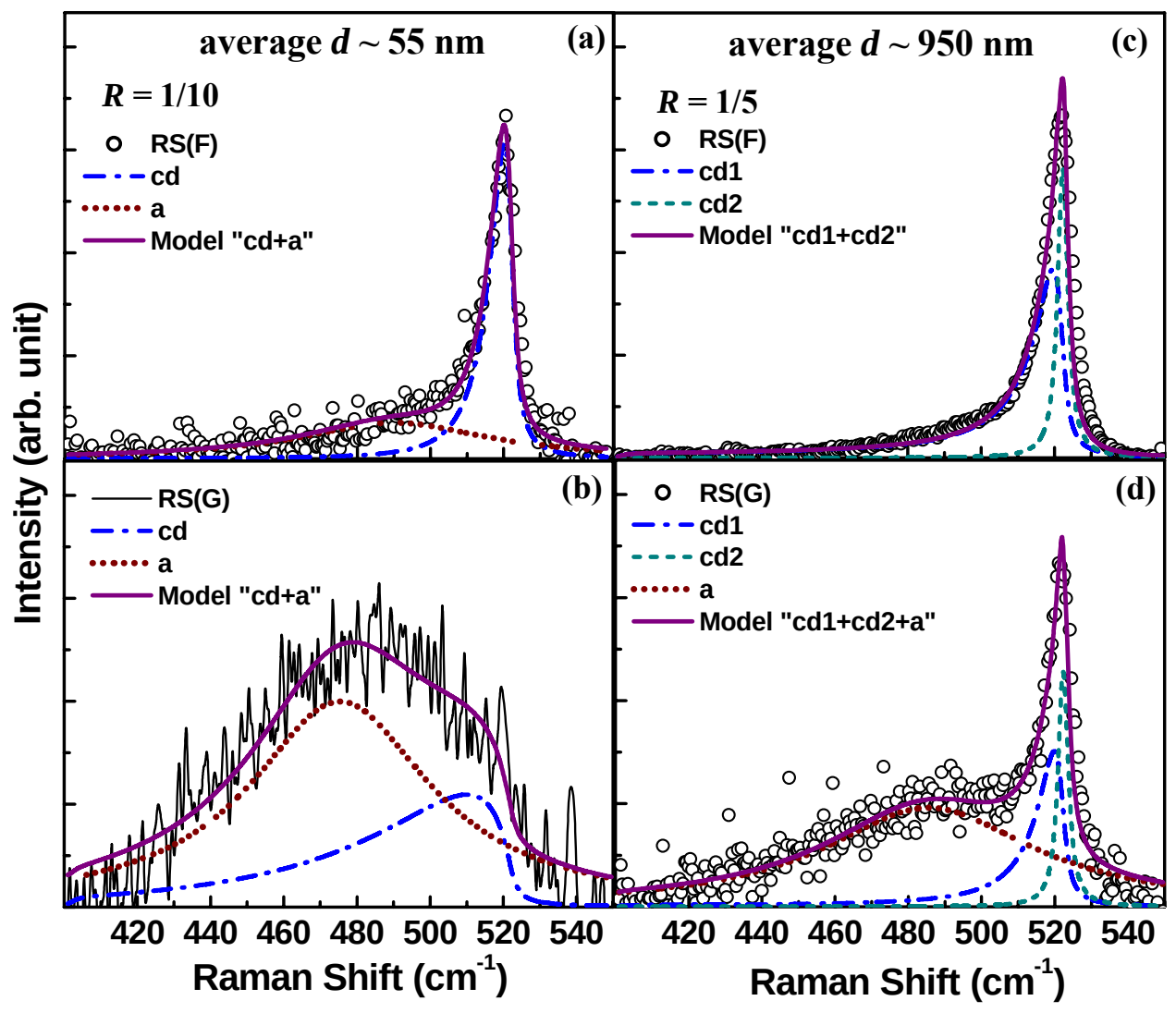

FIG. 4. Deconvolution results consisting of the total fit, and the deconvoluted peaks for the RS profiles of one samples from each thickness set of different $d$. The deconvolution of RS(F) and RS(G) profiles of $R$ = 1/10 film having $d \approx 55$ nm is shown in parts (a) and (b) respectively, while the deconvolution of RS(F) and RS(G) profiles of $R$ = 1/5 film having $d \approx 950$ nm is shown in parts (c) and (d) respectively. The respective fitting models are mentioned in each figure.

Table 1. Microstructural information such as mean crystallite sizes, their distribution [$\sigma$ (nm)] and fractional compositions obtained from deconvolution of bifacial RS profiles of the samples.

| Sample | Bifacial RS | Fitting Model | Small grain (cd1) | | Large grain (cd2) | | a-Si:H |
|---|---|---|---|---|---|---|---|
| | | | Size, [$\sigma$] (nm) | $X_{c1}$ (%) | Size, [$\sigma$] (nm) | $X_{c2}$ (%) | $X_a$ (%) |
| ***d*≈950nm** | | | | | | | |
| *R*=1/1 | RS(F) | cd1+cd2 | 6.4, [1.87] | 31 | 70.4, [0] | 69 | 0 |
| | RS(G) | cd1+cd2+a | 6.4, [1.52] | 16.4 | 97.6, [0] | 47.6 | 36 |
| *R*=1/5 | RS(F) | cd1+cd2 | 7.3, [1.98] | 29.5 | 73.8, [0.06] | 70.5 | 0 |
| | RS(G) | cd1+cd2+a | 7.7, [1.93] | 17.3 | 103.3, [0] | 50 | 32.7 |
| *R*=1/10 | RS(F) | cd1+cd2 | 8.4, [2.26] | 48.1 | 87, [0] | 51.9 | 0 |
| | RS(G) | cd1+cd2+a | 7.7, [1.59] | 23.4 | 101, [0] | 48 | 28.6 |
| ***d*≈55nm** | | | | | | | |
| *R*=1/1 | RS(F) | cd+a | 7.9, [1.93] | 44.7 | -- | 0 | 55.3 |
| | RS(G) | cd+a | 2.6, [0.64] | 29.6 | -- | 0 | 70.4 |
| *R*=1/10 | RS(F) | cd+a | 8.1, [1.65] | 71 | -- | 0 | 29 |
| | RS(G) | cd+a | 2.6, [0.50] | 32.7 | -- | 0 | 67.3 |

The microstructural composition of the films deduced from the deconvoluted RS data is presented in Table 1. The percentage volume fractions of the small and large crystallite grains and amorphous content are denoted by $X_{c1}$, $X_{c2}$ and $X_a$. The trend indicated by these values are in good agreement with the outcome of SE results, although a strict similarity between the absolute values deduced from these two different tools operating on different principles and at different length scales is neither possible, nor reasonable to aim for.

## 4. SUMMARY

Single phase nano (or synonymously, micro) crystalline silicon films were studied with different microstructural tools to elucidate the film microstructure at different stages of growth and for films deposited under different levels of $H_2$ dilution. Our results show the microstructural evolution with film growth for varying $H_2$ dilutions in the context of the crystallite grain growth, aggregation and variation in the percentage volume fractions of the constituent grains of different sizes. An initial increase in the $H_2$ dilution results in reduction of defects, better crystallinity and crystallite column formation. However, increasing the $H_2$ dilution further results in an increase in void fraction.

Our SE results demonstrate the presence of two distinct mean sizes of crystallites in the films after a certain growth stage, depending on the $H_2$ dilution level, which substantiates the rationale of taking into account a crystallite size distribution while analyzing microstructural data. The deconvolution of experimentally observed RS profiles of such nano and microcrystalline silicon films where a CSD is evident, has been done using a bimodal size distribution of large crystallite grains (≈70-80nm) and small crystallite grains (≈6-7nm), and the fractional compositional analysis of the films obtained by this methodology is strongly corroborative with the findings of SE and AFM studies.

## ACKNOWLEDGEMENTS

Financial support from Swiss National Science Foundation (#20720-109486) and Department of Science and Technology, New Delhi is gratefully acknowledged.